\documentclass[pre,floatfix,showkeys,twocolumn,a4paper,showpacs,superscriptaddress,nofootinbib]{revtex4-1}

\usepackage[utf8]{inputenc}
\usepackage[english]{babel}
\usepackage{amsmath,amsfonts,amssymb}
\usepackage[dvips]{color,graphicx,hyperref}

\begin{document}

\title{Communities and beyond: mesoscopic analysis of a large social network with complementary methods}

\author{Gergely Tib\'ely}
\affiliation{Institute of Physics and HAS-BME Cond.~Mat.~Group, BME, Budapest, Budafoki \'ut 8., H-1111}

\author{Lauri Kovanen}
\author{M\'arton Karsai}

\author{Kimmo Kaski}
\affiliation{BECS, Aalto University, P.O. Box 12200, FI-00076}

\author{J\'anos Kert\'esz}
\affiliation{Institute of Physics and HAS-BME Cond.~Mat.~Group, BME, Budapest, Budafoki \'ut 8., H-1111}
\affiliation{BECS, Aalto University, P.O. Box 12200, FI-00076}

\author{Jari Saram\"aki}
\affiliation{BECS, Aalto University, P.O. Box 12200, FI-00076}
\email{jari.saramaki@tkk.fi}

\date{\today}

\begin{abstract}
Community detection methods have so far been tested mostly on small empirical networks and on synthetic benchmarks. Much less is known about their performance on large real-world networks, which nonetheless are a significant target for application. We analyze the performance of three state-of-the-art community detection methods by using them to identify communities in a large social network constructed from mobile phone call records. We find that all methods detect communities that are meaningful in some respects but fall short in others, and that there often is a hierarchical relationship between communities detected by different methods. Our results suggest that community detection methods could be useful in studying the general mesoscale structure of networks, as opposed to only trying to identify dense structures.
\end{abstract}
\pacs{89.75.Fb,89.75.Hc,89.75.-k,89.65.-s}
\keywords{community detection; complex networks; social networks; mobile phone}
\maketitle

\section{Introduction}


Large complex networks have different levels of organization. On the microscopic level networks are composed of pairwise interactions, but it is the macroscopic level that has received most attention in recent years. We now know that diverse networks exhibit similarities for example in degree distribution, average path length, and clustering coefficient. While the structure is interesting in its own, it also has a significant influence on the dynamic processes taking place on the network, such as spreading, diffusion, and synchronization \cite{spreading,diff,sync}.

The intermediate mesoscopic scale has turned out be more elusive to describe. It is this scale where we can identify for example motifs \cite{UriAlon,intensity} and dense clusters of nodes commonly known as \emph{communities}. Although communities are relevant for understanding the structure of and the dynamics on networks, even their exact definition is still a controversial issue. Thus it comes as no surprise that the art of \emph{community detection} has grown into a swarming field of diverse methods \cite{SFreview}. Many features of real-world networks add to the complexity of the task. Real networks are often hierarchical and hence small communities may reside inside larger ones, communities may overlap if nodes participate in several communities, and even more complications arise if we take into account link weights that represent interaction intensity.

Until recently, the performance of community detection methods has mainly been tested on small empirical networks with typically no more than 100 nodes, which allows the evaluation of quality by visual inspection. However, several networks of considerable interest are much larger, often with $10^6$ nodes or more: data on WWW, mobile phone call records, electronic footprints of instant messaging users, and networks of social web such as Facebook etc. Only few methods are efficient enough to handle such networks \cite{AndreaLanch,SF_test,Louvain,SantoCharacterizing}---to be successful, a community detection method must be computationally efficient in addition to being accurate.

More systematic comparisons have been recently carried out using synthetic benchmark networks with built-in community structure \cite{SF_bechmark1, SF_bechmark2}. While benchmarks are useful in evaluating performance, even their authors acknowledge that they only represent the first step. No benchmark fully incorporates the spectrum of properties commonly observed in real-world networks. Some recent benchmarks do allow heterogeneous distributions for degrees and community sizes, but many other properties are still missing, such as high clustering, existence of cliques \cite{CP}, overlapping communities \cite{Ahn}, assortativity \cite{assort}, and the prevalence of motifs \cite{Alon}. This distorts the evaluation of algorithms that depend on (or benefit from) the existence of these features. For example, clique percolation has been successfully used on real-world networks \cite{CP,cpm_apps,biocpm,scicpm} but does not perform well on synthetic benchmarks---mainly due to its strict requirement for communities to consist of adjacent cliques \cite{SF_bechmark2}. 

In this paper we take three widely-applied methods, each based on a different underlying philosophy, and compare their performance on a large real-world social network constructed from mobile phone call records. Unlike with benchmark networks, we do not know the ``correct'' community structure of the network. Therefore, we introduce new measures that allow us to investigate the differences and similarities of the detected community structures.

The paper is organized as follows. Section \ref{sec:community_detection_methods} describes the choice of community detection methods and Section \ref{sec:data} introduces the data set. Section \ref{sec:results} presents the results of our analysis where we first analyse the properties and statistics of individual community structures and then turn to a pairwise comparison to quantify the differences between communities. Finally in Section \ref{sec:conclusions} we present conclusions.

\section{Choice of methods}
\label{sec:community_detection_methods}

As we intend to study a large network, the first selection criterion is only practical: methods with running time $O(N^2)$ or slower cannot be included. We use three methods that not only fill this requirement but in addition have performed well in previous comparisons or in practice: the \emph{Louvain method} (LV) \cite{Louvain}, the \emph{Infomap} (IM) \cite{IM} and the \emph{clique percolation} (CP) \cite{CP}. 

We consider an undirected network $G=(V,E)$, where $V$ is the set of $N$ nodes and $E$ the set of $L$ edges. The \emph{degree} $k_i$ is the number of neighbors node $i$ has, $k_i=|\{j|(i,j) \in E \}|$. For mathematical purposes a \emph{community} $c$ is simply a set of nodes, $c \subseteq V$, and we denote community size by $S=|c|$. The communities detected by one method constitute a \emph{community structure} $C = \{c_1,\ldots,c_{n_c}\}$. A \emph{partition} $P$ is a special community structure where each node belongs to exactly one community, i.e.\ $c_i \cap c_j = \emptyset$ if $i \neq j$ and $\bigcup_{i=1}^{n_c} c_i = V$.

All three methods can be extended to handle weighted networks where each edge has a numerical weight $w_{ij}$. In this paper we only consider positive weights; $w_{ij} = 0$ is equivalent to $(i,j) \notin E$. The weighted counterpart of degree is \emph{node strength}: $s_i = \sum_{(i,j) \in E} w_{ij}$.

The \emph{Louvain} method (LV) \cite{Louvain} was the best of the modularity optimization methods tested in \cite{SF_test}. Modularity is the expected value of the difference of the number of edges inside communities in the actual network and in a random network with the same degree sequence \cite{Newman_modularity}:
\begin{equation}
Q=\sum_{c \in P} \biggl[ \frac{L_c}{L} - \Bigl( \frac{d_c}{2L} \Bigr)^2 \biggr]
\label{eq:Q}
\end{equation}
where $L_c$ is the number of edges inside community $c$ and $d_c = \sum_{i\in c} k_i$ its total degree. In the weighted version all quantities are replaced by their weighted counterparts: $L_c$ by the total sum of weights inside a community, $d_c$ by the sum of node strengths and $L$ by the sum of edge weights in the whole network.

Because modularity optimization is an NP-complete problem \cite{NPcomplete_proof}, LV uses a greedy heuristic to find a local optimum. Each node is initially a separate community, i.e.\ $c_i = \{i\}$. Neighboring communities\footnote{Two communities are neighbors if there is at least one link between them.} are merged in random order so that modularity increases maximally at each step until a local maximum is reached. Resulting communities are then shrunk into ``super-nodes'' and the optimization is repeated on the new ``renormalized'' network. The two steps---optimization and renormalization---are repeated recursively until no further improvement of modularity is possible. 

The local heuristic of LV seems to avoid some of the resolution issues of modularity. In addition, the renormalized networks can be understood as different levels of a hierarchical community structure.

The \emph{Infomap method} (IM) \cite{IM} came out on top in a recent state-of-the-art benchmark comparison \cite{SF_test}. The idea is to describe a random walker with a two-level coding scheme: the higher level has a single codebook for communities, on the lower level each community has its own codebook with a special exit code for moving out of the current community. The optimal partition corresponds to the codebook with the minimum description length: too small communities increase the description length due to higher frequency of community crossings, while communities containing too many nodes require longer description. In weighted networks the random walks are biased towards edges with higher weight. Since an exhaustive search for the optimal partition is not feasible, Infomap employs a heuristic similar to the one used in LV.

\emph{Clique percolation} (CP) \cite{CP} has been successfully applied to large empirical graphs, e.g. to study the dynamics of social groups \cite{cpm_apps}. A $k$-clique is a fully connected subgraph of $k$ nodes, and two $k$-cliques are considered adjacent if they share $k-1$ nodes. As the name suggests, clique percolation defines communities as connected $k$-clique components: a CP community is a maximal set of $k$-cliques such that there is a path of adjacent $k$-cliques between them. Different values of $k$ yield different community structures, and communities obtained with a larger value of $k$ reside inside those obtained with a smaller value. To select the best value of $k$ we follow Ref.\ \cite{CP} and use the smallest value for which there is no giant percolating community.

There are significant differences between CP and the other two methods. Both LV and IM use a stochastic optimization scheme while CP is entirely deterministic. In addition, LV and IM yield a partition but CP does not. With CP the nodes that do not belong to any $k$-clique are left outside communities, and if a node belongs to several $k$-cliques it may belong to more than one \emph{overlapping} community. The fact that CP does not provide a partition is not necessarily a bad thing: sparse regions of the network do not appear as communities, and e.g. in social networks individuals often do belong to multiple groups, such as family, friends, and colleagues.

To define the weighted clique percolation (wCP) \cite{wCP} we need the concept of clique \emph{intensity}, defined as the geometric mean of edge weights. In wCP we use a value of $k$ that would give a giant community in the unweighted case, but only include those $k$-cliques that have intensity larger than some predefined threshold $I_>$. Analogously to the unweighted case, $I_>$ is set to the largest value for which there is no giant community.

Notes on applying the methods are given in Appendix \ref{appendix:notes_on_application}.

\section{The data}
\label{sec:data}

Our empirical test network is a mobile phone call network constructed from billing records of seven million customers of a single mobile phone operator whose customer base covers about $20\%$ of the population in its country. The records cover a period of 126 days. To ensure anonymity of customers, phone numbers have been replaced by surrogate keys. Data from the same operator has been previously studied in \cite{HumanMobility, cpm_apps, StrongTies}.

For this study we use only voice calls, and only those that take place between customers of the operator in question. In addition we exclude edges where only one person has made calls to the other during the whole period. We study only the largest connected component which has $N=4.9\times 10^6$ nodes and $L=10.9\times 10^6$ edges (mean degree $\langle k \rangle \approx 4.44$).\footnote{The largest connected component contains 92 \% of nodes and 98 \% of edges; the second-largest component has only $47$ nodes.} 

The edge weights in the weighted network are defined as sums of call durations (in seconds) between the two customers. The average weight is $\langle w \rangle \approx 4634$ seconds.

Using a large social network enables us to relate the findings to known characteristics of such networks \cite{StrongTies}. It is known that the overlap of local neighborhoods of adjacent nodes increases with edge weight\footnote{Except for the very largest edge weights, where the relation is reversed.} \cite{Onnela}, as conjectured in the ``weak ties'' hypothesis of Granovetter \cite{Granovetter1}. This feature should be reflected in correlations between edge weights and communities. We can also study structural features of communities and evaluate whether they represent meaningful social communities.


\section{Results}
\label{sec:results}

We analyse \emph{single} community structures detected by each method. Both LV and IM are stochastic methods and therefore give a slightly different partition on every run; however, as shown in Appendix \ref{appendix:stability} the qualitative properties of the communities are stable enough to justify the comparison.

Appendix \ref{appendix:notes_on_application} contains detailed notes about the application of the three methods. In brief, we use parameters $k=3$ for CP and $k=4$ with $I_> = 3093$ for wCP---these are the only two methods with explicit parameters---and with LV we only study the first level of the hierarchical community structure since other levels yield communities that are implausibly large in the social context.

\subsection{Community size distributions}
\label{sec:size-distribution}

\begin{figure*}[t]
  \includegraphics[width=160mm]{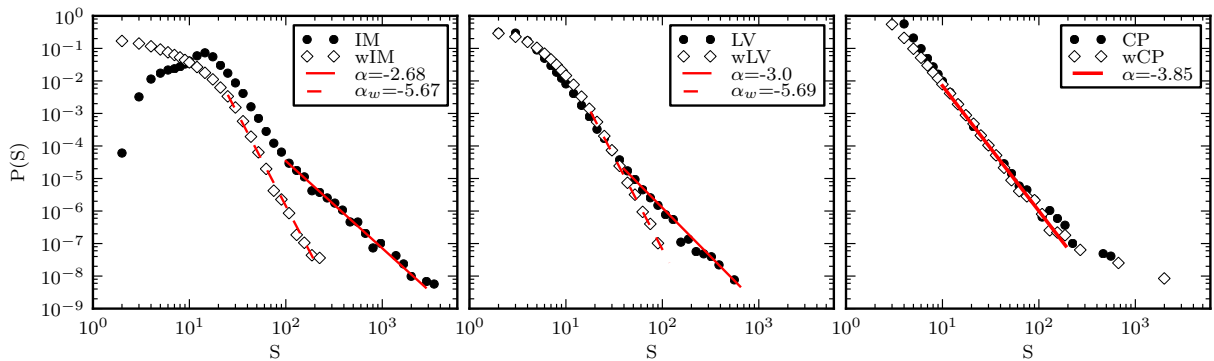}
  \caption{\label{fig:sizedistributions}(Color online) Community size distributions for IM, LV and CP and their weighted versions. The parameter $\alpha$ denotes the exponent when the tails are fitted a  power-law distribution $P(S) \propto S^{\alpha}$; solid lines correspond to the unweighted  $\alpha$ and dashed lines to
  the weighted  $\alpha_w$. }
\end{figure*}

Figure \ref{fig:sizedistributions} shows the community size distributions for all methods. All distributions are broad, as suggested by previous results \cite{Clauset2004,CP, SantoCharacterizing}.

For IM, the tail of the size distribution appears power-law-like. Very small communities are rare.
The community structure of wIM is notably different. The weighted communities are smaller, and the distribution is now monotonously decreasing.

Even though the largest LV communities are an order of magnitude smaller than in IM, LV still produces larger communities than its weighted variant wLV. Both LV and wLV have monotonous community size distributions, and small communities are more prevalent than in IM. The power-law exponents for the tails are similar when comparing LV to IM and wLV to wIM.

For CP and wCP the size distributions are well approximated by a power law. This is expected, as the communities are detected close to the critical point where a giant community would emerge. The largest deviation from power law behaviour is in the tail. The largest wCP communities are larger than those in CP because 3-cliques are used for wCP and 4-cliques for CP. Although these communities partially overlap (see Section \ref{sec:overlap}), the 3-clique communities extend far beyond the 4-clique communities.


\subsection{Visual observation of small communities}
\label{sec:visual_observation}

\begin{figure*}[ht]
  \mbox{\includegraphics[width=0.5\textwidth]{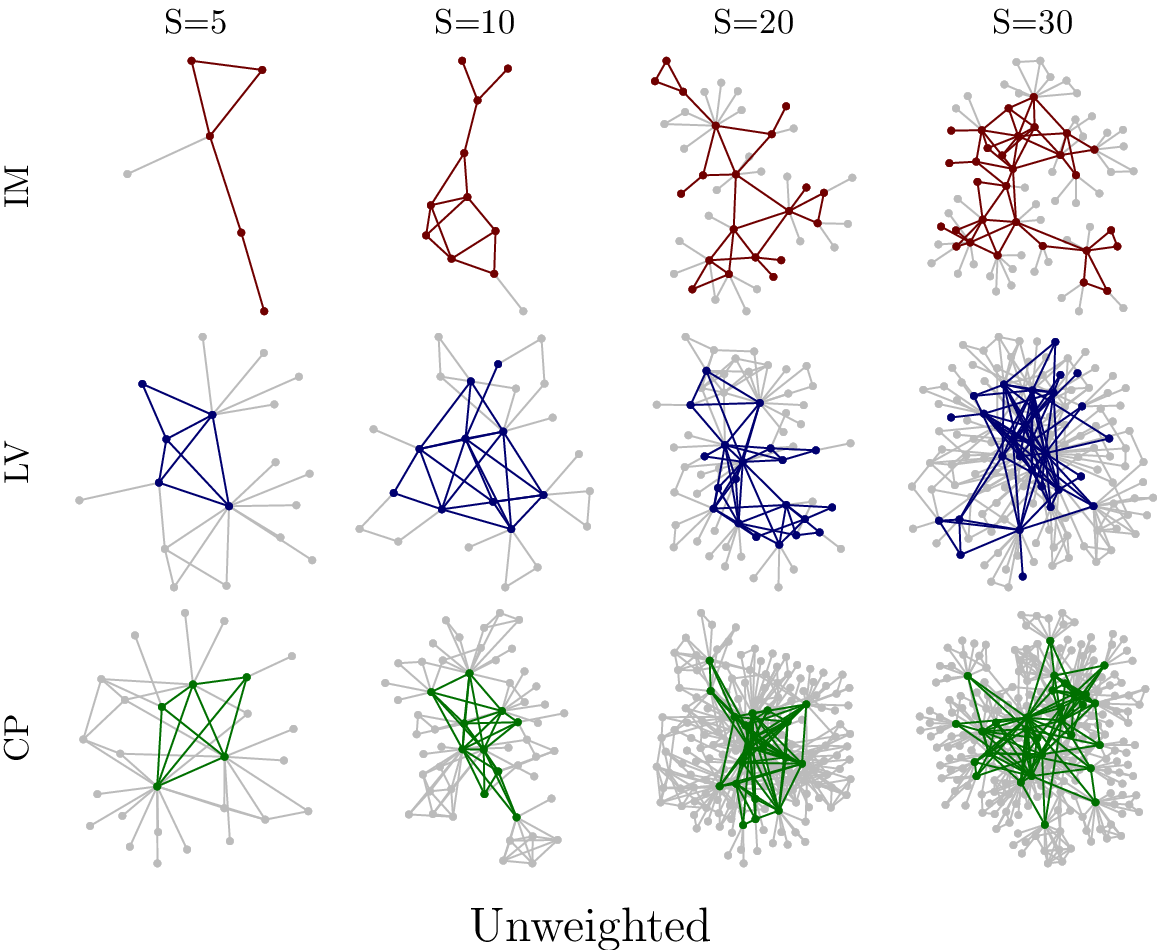}
    \rule{0.4pt}{67mm}
    \includegraphics[width=0.5\textwidth]{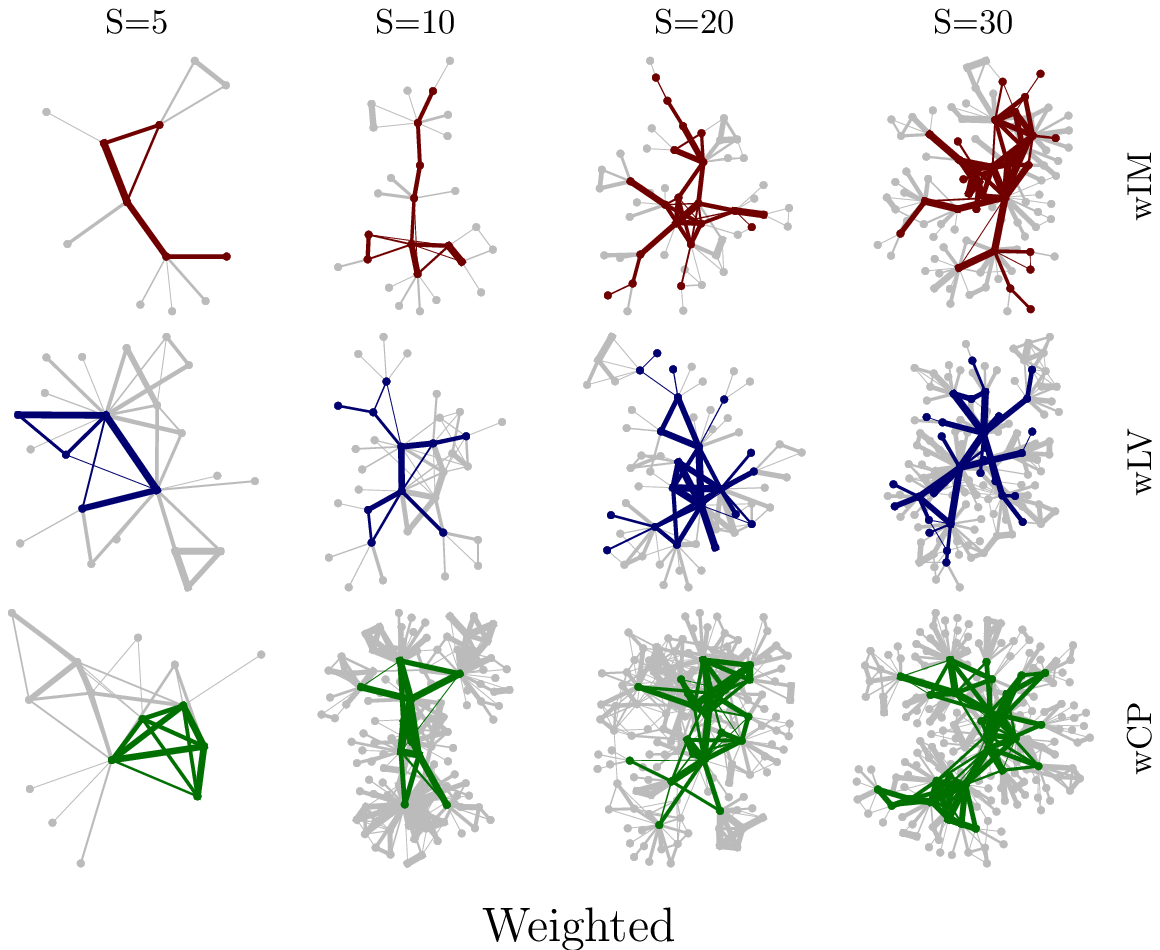}}
  \caption{\label{fig:community_examples}(Color online) Typical (left) unweighted and (right) weighted communities of different size. These communities have been manually selected from a large random sample of communities with the intention of portraying archetypal examples. Colored (dark gray) nodes and edges denote nodes inside a single community, and the light gray nodes are the first neighbors of the nodes in the community. In weighted communities the edge width in is proportional to the logarithm of edge weight, with the restriction that edges with $w_{ij} \leq 300$ (5 min) have the minimum width and those with $w_{ij} \geq 14400$ (4 h) the maximum width.}
\end{figure*}

The qualitative properties of small communities can be estimated visually, similarly to evaluating performance on small empirical networks. Fig.\ \ref{fig:community_examples} shows archetypal communities with $S=5$, $10$, $20$, and $30$, and their immediate network surroundings. Communities larger than this tend to be too complex to visualize in two dimensions.

Of all unweighted methods the CP communities are the least surprising: larger communities naturally appear only in dense parts of the network. Small LV communities consist of interconnected cliques, which coincides well with the general idea of social groups. The smallest IM communities with $S \leq 10$, however, are typically treelike and located at the ``edge'' of the network -- these communities are attached to the rest of the network by only few links. LV covers these sparse parts of the network with much smaller communities (see Fig.\ \ref{fig:tiling_example}).

When the weights are taken into account, the partition-based methods wIM and wLV tend to produce even more treelike communities that have the appearance of local ``backbones'' of the network. This is a natural consequence of the way wIM and wLV use edge weights; however, communities like these do not coincide well with the idea of dense social groups.

\subsection{Community density distribution}

\begin{figure}
  \includegraphics[width=0.90\columnwidth]{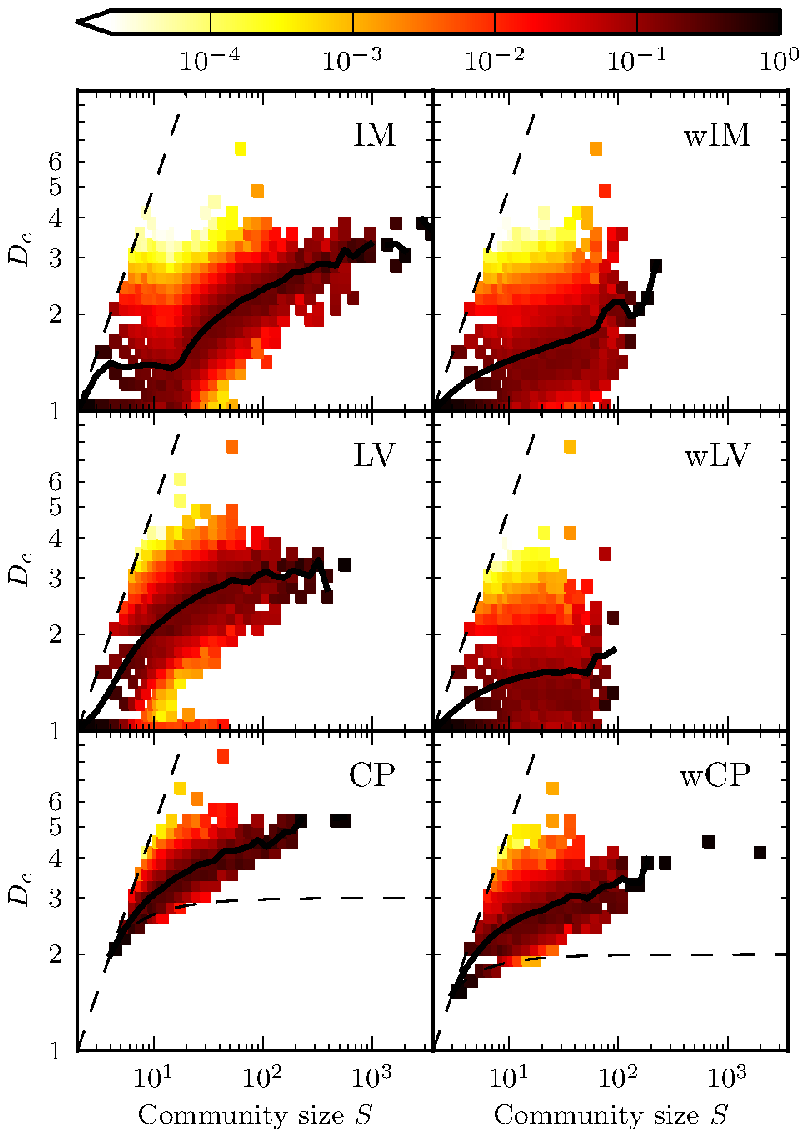}
  \caption{\label{fig:density}(Color online) The distribution of relative density $D_c = L_c/(S-1)$ for communities from each method. In all plots, each column represents a distribution and is normalized to one, the colors indicating probability density so that the darker the color, the higher the density (see color bar). The thick solid line     denotes the average value. The dashed straight line corresponds to cliques, for which $D_c = S/2$. For IM and LV the smallest density is 1, which corresponds to trees. For CP, the smallest possible density is indicated by the curved dashed line (see text).}
\end{figure}

Since some small communities were already observed to be treelike, we turn to more quantitative characterization of community density. Graph density is normally defined as the proportion of edges out of all possible edges, $L_c/\left[\frac{1}{2}S(S-1)\right]$. However, since communities are necessarily connected it is more illustrative to study density relative to the sparsest possible community, a tree with $S-1$ edges, as also done in \cite{SantoCharacterizing}: we define density as $D_c ={L_c}/\left({S-1}\right)$. In general $1 \leq D_{c} \leq S/2$ where the lower bound corresponds to trees and the upper bound to cliques. CP however doesn't allow trees; instead, the smallest possible density is reached when each new node adds only $k-1$ edges. In this case $L_c = \binom{k}{2} + (k-1)(S-k)$ which gives $D_{c} \geq {(k-1)(S-\frac{k}{2})} / \left({S-1}\right)$. For $S \gg k$ this is approximately $k-1$.

Figure \ref{fig:density} shows the distributions and average values of $D_c$ as function of community size. As expected, CP yields the densest communities. For IM the value of $D_c$ stays close to 1 until $ \approx 20$, which confirms the observation on the prevalence of small treelike communities. For LV the distribution has a curious bimodal shape in the range $20 < S < 50$: typical LV communities of this size have $D_c$ from 2 to 4, but there is a small number of LV communities that are trees ($D_c = 1$) but none that are almost trees. A closer inspection (not shown) of these trees reveals that they are stars.

The plots for weighted communities in Fig.\ \ref{fig:density} suggest that weights make the communities more similar across methods. Both wIM and wLV communities are more treelike, as already seen in Section \ref{sec:visual_observation}.

Treelike communities do not fit well either with the idea of social groups, or that of communities in general being dense groups of nodes. However, if a network contains treelike regions, partition-based methods will correspondingly yield treelike communities\footnote{It has been shown that if there are nodes with a single link, for   modularity optimization they should always belong to the community   of the node to which they are connected~\cite{ArenasNJP}. By construction, this holds for IM as well.}, as also seen in Ref.\ \cite{SantoCharacterizing}. The abundance of treelike parts may just be a sampling artifact, as our network does not cover the whole population. Nevertheless, empirical data is rarely perfect, and a good community detection method should deal with this in a sensible way. One could argue that in treelike regions the network is so sparse that there isn't enough information about community structure. This makes CP's requirement---that nodes must participate in at least one clique to be assigned a community---appear meaningful. On the other hand, CP may yield communities where \emph{cliques} are arranged as chains or starlike patterns, which again does not coincide well with the idea of social groups. Fig.\ \ref{fig:density} indicates that in CP and wCP there are indeed some communities with densities close to the lower bound.

Whatever the interpretation, the detected treelike structures do provide information about the mesoscopic structure of the network. In other networks starlike structures can represent meaningful communities: for example in air transport networks the peripheral airports are connected to local hubs \cite{GuimeraAirlinePNAS}.

\subsection{Intra- and intercommunity edges}
\label{sec:edges}

\begin{figure}
  \includegraphics[width=0.90\columnwidth]{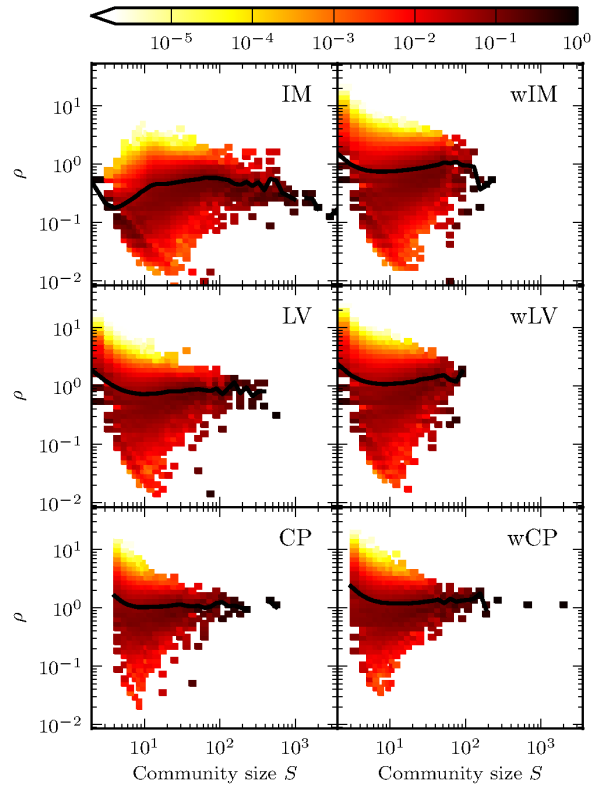}
  \caption{\label{fig:community_edges}(Color online) The distribution of $\rho(c)$ (Eq.\ \ref{eq:rho}) as function of community size for each method. The distributions are presented as in Fig.\ \ref{fig:density}, with a similar shading scheme. The black line denotes average value.}
\end{figure}

\begin{table}
  \caption{\label{table:community_weights}Edge weights inside and between communities. $\langle w\rangle$ denotes the average edge weight in the whole network, $\langle w_c\rangle$ the average weight for edges inside communities and $\langle w_{c-c}\rangle$ between communities. CP also has non-community nodes; $\langle w_{c-n}\rangle$ denotes the average weight between community and non-community nodes and $\langle  w_{n-n}\rangle$ between two non-community nodes.}
  \begin{tabular}{c|cccc}
    & $\langle w_{c} \rangle / \langle w \rangle$ & $\langle w_{c-c}             \rangle / \langle w \rangle$ & $\langle w_{c-n} \rangle / \langle w             \rangle$ & $\langle w_{n-n} \rangle / \langle w \rangle$  \\
    \hline
    IM & 1.14 & 0.69 &  & \\
    LV & 1.20 & 0.78 &  & \\
    CP & 1.20 & 0.57& 0.80 & 1.06\\
     \hline
     wIM & 1.65 & 0.18 & & \\
     wLV & 1.92 & 0.25 & & \\
     wCP & 2.57 & 0.43 & 0.57 & 0.73\\
  \end{tabular}
\end{table}

If the detected partitions are any good, nodes should have more edges to other nodes in the same community than to those in other communities. To measure this we define $\rho(c)$ as the ratio of total out- and in-degree of a community:
\begin{equation}
  \rho(c) = \frac{\sum_{i \in c} k_i^{\text{out}}}{\sum_{i \in c} k_i^{\text{in}}}=\frac{1}{2L_c}\sum_{i \in c} k_i^{\text{out}}~~~.
  \label{eq:rho}
\end{equation}
Figure \ref{fig:community_edges} shows the distribution of $\rho(c)$ as function of community size.
With respect to this measure IM produces the most clear-cut communities: majority of IM communities have $\rho$ below one. The values for small communities are especially low, confirming the earlier observation that small IM communities are on the ``edges'' of the network. LV communities also have $\rho < 1$ on average, except for the smallest communities, but the values are not as low as with IM. Including weights increases the average value of $\rho$. wLV communities in fact have on average more links going outside the community than inside.

Because CP allows nodes to belong to multiple communities, a good community need not have a low value of $\rho(c)$. Also note that with CP a large fraction of edges are attached to non-community nodes. For CP (wCP) only 21.4 \% (18.6 \%) of edges and 21.8\% (25.4 \%) of nodes are inside communities; 47.6 \% (43.0 \%) of edges are between non-community nodes.

\begin{figure}
  \includegraphics[width=0.7\linewidth]{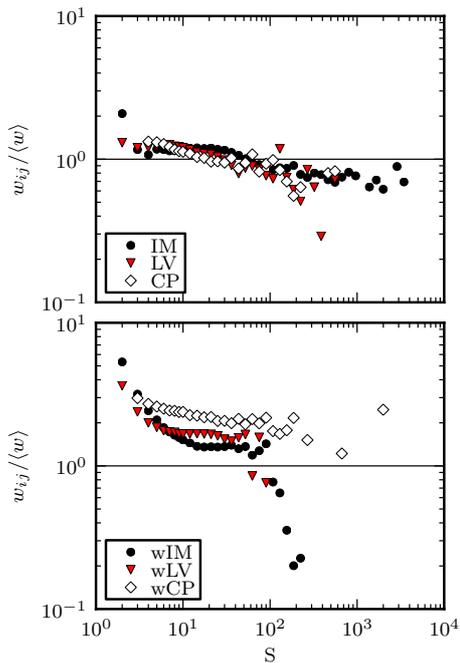}
  \caption{\label{fig:weights_in_communities}(Color online) Average edge weights     $w_{ij}/\langle w \rangle$ inside communities as a function of     community size $S$, normalized by the network average.}
\end{figure}

From earlier studies of mobile phone call networks \cite{StrongTies,Onnela} we know that there is a correlation between edge weight and neighborhood overlap, in agreement with the Granovetter hypothesis \cite{Granovetter1}. As nodes inside communities have overlapping neigbourhoods, we expect the links between communities to be on average weaker than those within communities. Table \ref{table:community_weights} shows that with all methods this is indeed the case. With weighted methods this result is of course not as surprising since weights were used in identifying the communities.

To see beyond averages, Fig.\ \ref{fig:weights_in_communities} displays the normalized average edge weight inside communities as function of community size. Most notably the edge weights in the largest communities are below the network average---even for wIM and wLV.

\subsection{Neighbourhood overlap}
\label{sec:overlap}

\emph{Neighbourhood overlap} quantifies the similarity of a node's neighbourhood in two community structures. If $\mathcal{N}_i(C_j)$ is the set of those neighbours of node $i$ that belong to its community in $C_j$, the neighborhood overlap is defined as Jaccard index of $\mathcal{N}_i(C_1)$ and $\mathcal{N}_i(C_2)$:
\begin{equation}
 O_i(C_1,C_2)=\dfrac{|\mathcal{N}_i(C_1) \cap \mathcal{N}_i(C_2)|}{|\mathcal{N}_i(C_1) \cup \mathcal{N}_i(C_2)|}~~~.
\end{equation}
Thus $O_i=1$ if the same neighbours of $i$ belong to its own community in both methods and $O_i=0$ if the sets do not overlap. In the case of CP we only consider nodes that participate in at least one community; for nodes that participate in several, we assign the node to the community where most of its neighbours reside.

\begin{figure}
  \includegraphics[width=0.7\linewidth]{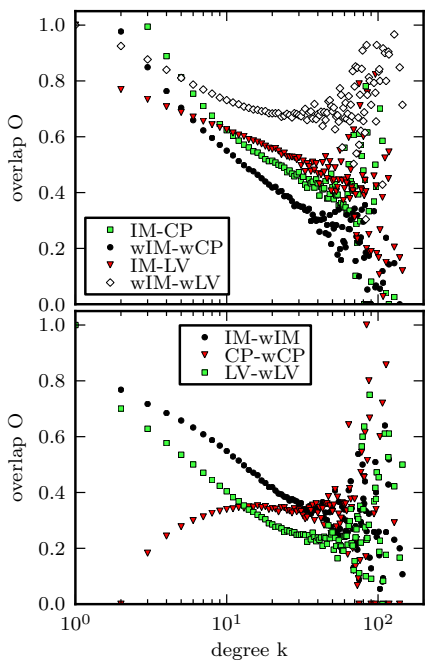}
\caption{\label{fig:overlap_pairs}(Color online) Average neighbourhood community overlap $O$ as a function of node degree $k$, between different methods (top) and unweighted and weighted versions of the same method (bottom).
}
\end{figure}

Figure \ref{fig:overlap_pairs} displays the average neighbourhood overlap as function of degree for selected method pairs\footnote{Instead of showing the results for all 15 method pairs we only present the most interesting cases.}. Nearly all pairs show a decreasing trend and thus in general community neighbourhoods of low-degree nodes are more similar. The IM--CP and wIM--wCP overlaps decrease the fastest, as the underlying philosophies are different and the large number of nodes not appearing in any CP community reduces the overlap. wIM and wLV show a better match than their unweighted counterparts, suggesting a similar and fairly strong response to edge weights. On the other hand, overlaps for IM--wIM and LV--wLV become small for large $k$, which suggests that taking weights into account considerably changes the partitions for these methods. With CP--wCP the opposite behaviour occurs because wCP is based on 3-cliques and many nodes that are included in a 3-clique are not included in any 4-clique.

\subsection{Nested communities}\label{sec:nested}

The above analysis shows that the three methods do not detect the same communities. It is however possible that they only detect different levels of a hierarchical community structure. If this is true, then the communities from one method should be the subset of another.

\begin{figure}
\includegraphics[width=0.58\columnwidth]{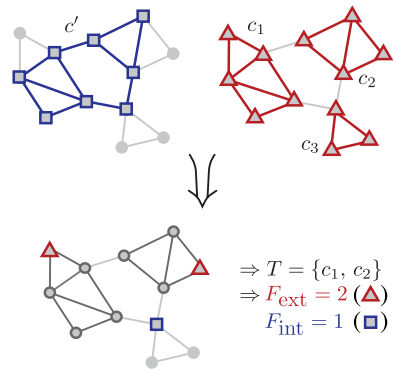}
\caption{\label{fig:tiling_illustration}(Color online) Illustration of tiling imperfection. The 8 nodes in $c'$ are spread over three different communities in another partition. Using $T = \{c_1,c_2\}$ gives the best tiling; including $c_3$ would reduce $F_{\text{int}}$ to 0 but increase $F_{\text{ext}}$ by 2. The value of tiling imperfection is $\mathcal{I} = 3/8$.}
\end{figure}

To address this question quantitatively we calculate how accurately a single community $c' \in P_i$ can be \emph{tiled} by the communities of another partition $P_j$. The best tiling is reached with set $T \subseteq P_j$ that minimizes the sum of \emph{external faults}
\begin{equation}
  F_{\text{ext}}(c',T)  = \sum_{c_j \in T} |c_j| - |c' \cap c_j|
\end{equation}
which equals the number of nodes in $T$ but outside $c'$, and \emph{internal faults}
\begin{equation}
  F_{\text{int}}(c',T) = |c'| - \sum_{c_j \in T} |c' \cap c_j|
\end{equation}
which equals the number of nodes in $c'$ but outside $T$. As illustrated in Fig.\ \ref{fig:tiling_illustration}, the minimum of $F_{\text{ext}}+F_{\text{int}}$ is reached when $T$ contains only those communities for which $|c' \cap c_j| > \frac{1}{2}|c_j|$, i.e.\ those $c_j \in P_j$ that share at least half of their nodes with $c'$. To allow comparing communities of different size we define \emph{tiling imperfection} $\mathcal{I}(c', P_j)$ as the ratio of this minimum total fault and community size:
\begin{equation}
\mathcal{I}(c',P_j)=\frac{\min(F_{\text{ext}}+F_{\text{int}})}{|c'|}~~~.
\label{eq:tiling_imperfection}
\end{equation}
Note that the aim of this measure is to quantify the subset-superset relationships of communities, which cannot be done with symmetric measures such as mutual information.

It is possible to generalize this measure also for general community structures,\footnote{If $T^* = \cup_{j \in T} c_j$, the generalized tiling is defined by $F_{\text{ext}}(c',T) = |T^*| - | T^* \cap c' |$ and $F_{\text{int}}(c',T) = |c'| - | T^* \cap c'|$. The optimal $T$ can now be constructed by first including (as before) the communities that share at least half of their nodes with $c'$, but then adding also those communities that contain more uncovered nodes of $c'$ (i.e. those in $c' \backslash T^*$) than new nodes outside $c'$. Here we however use the same definition of $T$ as for partititions to make the values more comparable.} such as the one produced by CP, but this is not advisable: if $c'$ would have nodes that are not included in any community of $C_j$, these nodes would automatically be internal faults and the tiling imperfection would be misleadingly high. To correct for this we define \emph{inclusion imperfection} $\mathcal{I}^{\ast}(c',C_j)$ similar to tiling imperfection but nodes may be counted as internal faults only if they are covered by both community structures.

\begin{figure}[t]
  \includegraphics[width=0.7\linewidth]{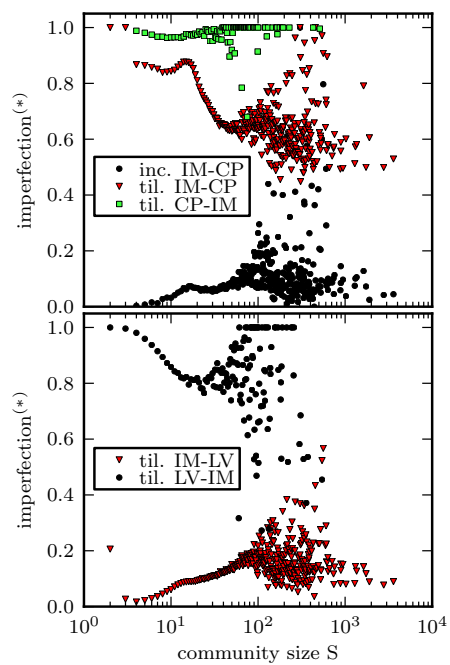}
\caption{\label{fig:tiling_plots}(Color online) (Top) Tiling imperfection $\mathcal{I}$ and inclusion imperfection $\mathcal{I}^{\ast}$ between IM and CP. 
(Bottom) Tiling imperfection $\mathcal{I}$ between IM and LV. 
}
\end{figure}

Results for tiling measures are shown in Figure \ref{fig:tiling_plots}. Comparing the tiling and inclusion imperfections for IM--CP, especially for small communities, illustrates the difference of these two measures: \emph{tiling} imperfection is high since small IM communities are treelike and therefore not included in any CP community; low values of \emph{inclusion} imperfection, however, show that CP communities tend to be subsets of IM communities. High values of CP--IM tiling imperfection shows that the reverse is not true.

\begin{figure}[t]
\includegraphics[width=0.8\columnwidth]{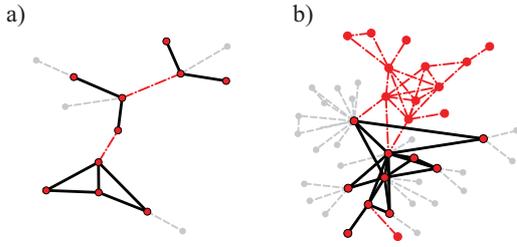}
\caption{\label{fig:tiling_example} (Color online) Typical cases of tiling with IM (red or dark gray) and LV (black) communities of size $S=10$. Light gray nodes are the first neighbors of the community to be tiled. \textbf{(a)} Example of perfect tiling $\mathcal{I}=0$ when IM community (red nodes) is tiled with LV communities (black edges). A typical IM community with $S=10$ is located in a treelike region of the network, and LV covers such regions with very small communities. \textbf{(b)} Example of tiling imperfection $\mathcal{I}=1$ when LV community (black edges) is tiled with IM communities (in red). A typical LV community with $S=10$ is in a somewhat denser part of the network, where the IM communities are much larger.}
\end{figure}

The low tiling imperfection for IM--LV and high for LV--IM shows that IM communities tend to be supersets of LV communities. The extreme values for small communities indicate that nearly all small IM communities can be perfectly tiled with LV communities, while small LV communities can almost never be tiled with IM communities.\footnote{Note that $\mathcal{I}$ may only take values that are fractions of community size: e.g. with $S=5$ the smallest non-zero value is $0.2$, and to get an \emph{average} value of $O(10^{-2})$ the vast majority of IM communities must have $\mathcal{I} = 0$.} A typical tiling of small IM and LV communities is shown in Fig.\ \ref{fig:tiling_example}.

\section{Conclusions and discussion}
\label{sec:conclusions}

Benchmarks are helpful if the methods are to be tested for sensitivity to particular properties, such as hierarchical structure or broad distribution of community sizes. Real-world networks, however, are incomparably more complicated, often inhomogeneous in many respects and usually contain many different kinds of mesoscopic structures. Good performance on benchmark graphs does not assure that communities identified in real data are meaningful. Our analysis of the Infomap, Louvain and clique percolation methods applied to a large social network reveals that while all the three methods do detect reasonable communities in some respects, they still come short in others.
 
With all these methods the edge weights were higher inside communities than between them, in accordance with the Granovetter hypothesis \cite{Granovetter1}; distributions of community sizes were broad, as expected; and tiling imperfection revealed that while IM and LV produce different partitions, they have a hierarchical relation where LV communities tend to be inside IM communities. On the other hand, both IM and LV yield treelike communities which does not coincide well with the notion of a social community, and using edge weights makes the communities even sparser. In contrast, CP clusters are always found in dense regions of the graph and are therefore often meaningful; as a downside CP may end up discarding some important parts of communities. 

A natural question is how well our findings can be generalized to other types of networks. Analysis of multiple datasets is beyond the scope of this work, but some speculation can be done. Broad community size distributions have already been observed in a number of studies \cite{SantoCharacterizing, CP, AndreaLanch}. Considering the numerous treelike communities, similar sparse regions occur in other networks as well. For example, \cite{SantoCharacterizing} found that the density of communities can vary widely across different network types; e.g. the Internet has very sparse communities while information networks (like arXiv citations) have dense ones. The similarity of IM and LV may hold too because both partition the network and their heuristics are similar. \cite{SantoCharacterizing} observed that two very different partitioning methods resulted in similar communities in terms of statistical properties. On the other hand, the difference between CP and the partition-based methods is likely to manifest itself for various networks.

In large sparse networks partitioning methods inevitably identify some questionable regions as communities. The trees, starlike formations and stars detected by IM and LV do, however, bear mesoscopic structural meaning: they too are building blocks of the network. The same topological structure may be considered a community for one purpose but not for some other---a star is hardly a social community but may reasonably be considered as one in for example biochemical networks \cite{SantoCharacterizing}.

It would seem that the analysis of large empirical networks would benefit from the use of complementary community detection methods and a comparison of the identified structural features. Instead of just devising ever more efficient community detection methods it might be more beneficial to take into consideration the existence of different types of mesoscopic structures, as opposed to fixating on a predefined idea of dense communities.

\begin{acknowledgments}
The project ICTeCollective acknowledges the financial support of the Future and Emerging Technologies (FET) programme within the Seventh Framework Programme for Research of the European Commission, under FET-Open grant number: 238597. We acknowledge support by the Academy of Finland, the Finnish Center of Excellence program 2006-2011, project no.\ 129670. JK thanks OTKA K60456 and TEKES for partial support. We thank Albert-L\'aszl\'o Barab\'asi for the data used in this research.
\end{acknowledgments}

\appendix

\section{Notes on applying the methods}
\label{appendix:notes_on_application}

\emph{The Louvain method}. The LV agglomeratively builds larger communities until no improvement in modularity can be achieved. Our data yielded very large communities with sizes up to $S\simeq 5\times10^5$ nodes both for LV and wLV, and hence we adopted the view that the different renormalization levels correspond to different levels of hierarchical organization\footnote{Note,  however, that this assumption has not yet been verified e.g. with benchmarks.}, as suggested in Ref.\ \cite{Louvain}. To obtain meaningful, smaller social communities and to be able to compare results with other methods we chose to use the first level, i.e.\ before the first merger of communities was made. This step revealed another feature of LV: while the modularity value is quite similar regardless of the order the nodes are processed in, the size of the largest community varies greatly. We use a partition where the size of the largest community is around $10^3$ since this makes sense in the social context. Because LV uses a local heuristic and we are dealing with a very large network, it is reasonable to assume that the statistical properties of the partitions are on average similar and do not vary as much as the size of the largest community. For a detailed description of the stability of both LV and IM, see Appendix \ref{appendix:stability}. In addition the LV algorithm can in some cases produce \emph{disconnected} communities. Only few such communities were encountered, and we dealt with this by turning each connected component into a community. Code for the algorithm is available for download~\cite{LV_code}.

\emph{The Infomap method}. The implementation code for Infomap is available for download \cite{IM_code}. No changes to the code were required.

\emph{Clique percolation}. For CP we need to select the value of $k$ such that there is no percolating cluster. For our data, $k=3$ gives rise to a giant community but $k=4$ does not and thus we select $k=4$.

\begin{figure}
  \includegraphics[width=0.8\columnwidth]{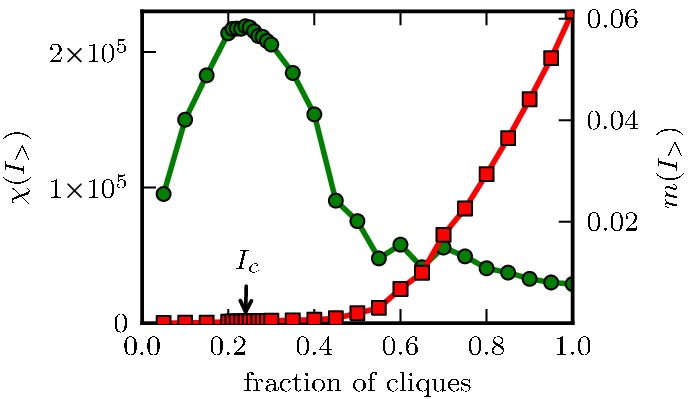}
  \caption{\label{fig:CP_percolation_point} (Color online) To find the critical     threshold $I_c$ for wCP we build up communities by adding  cliques in descending order of intensity $I$, and monitor the  largest component size $m(I_>)$ ($\square$) and susceptibility $\chi(I_>)$ ($\circ$). The     transition occurs when about 24 \% of cliques have been added ($I_> \approx 3093$).}
\end{figure}

For the weighted wCP we start with $k=3$ and find the threshold intensity $I_>$ for which the giant community disappears.\footnote{Note that with $k=4$ the weighted communities  would be identical to the unweighted ones, as in the absence of  percolation the intensity threshold would be set to 0. Using $k=2$ on the other hand would correspond to simply using a weight  threshold on single edges.} Thus we look for the percolation point using clique intensity as the control parameter \cite{wCP} and set the intensity threshold $I_>$ slightly below the critical point.  This point can be identified by the maximum of the susceptibility-like quantity
\begin{equation}
  \chi=\sum_{S_{\alpha}\neq             S_{max}}{S_{\alpha}^2}/(\sum_{\beta}{S_{\beta}})^2 
  \label{suscept}
\end{equation}
where $S$ is community size, and $\alpha$ and $\beta$ index the communities. We varied  $I_>$ while monitoring the order parameter $m(I_>)$ and the susceptibility $\chi(I_>)$ (see Figure \ref{fig:CP_percolation_point}). When 24 \% of the cliques have been added in order of descending intensity, a giant cluster emerges, while susceptibility shows a pronounced peak. This point corresponds to the critical intensity $I_c \approx 3093$, which was chosen as our threshold. For CP and wCP, we applied the fast algorithm introduced in \cite{FastCP}. A sample implementation can be found at \cite{FastCP_link}

The running times of all the algorithms used are displayed in Table \ref{table:runtimes}. LV and CP are extremely fast, while Infomap takes a few days to complete. All runs were done on a standard desktop machine, utilizing a single processor.


\begin{table}
  \caption{\label{table:runtimes}Running times of the different             algorithms on our data set of $N=4.9\times 10^6$ nodes and             $L=10.9\times 10^6$ links.}
  \begin{tabular}{r|cc}
    & unweighted & weighted \\
    \hline
    Louvain & 2 min 7 s & 1 min 30 s \\
    Infomap & 46 h 44 min & 3 h 20 min \\
    Clique percolation & 2 min 10 s & 4 min 52 s \\
  \end{tabular}
\end{table}

\section{Stability of the stochastic methods} \label{appendix:stability}

Both IM and LV are stochastic methods, and therefore the partitions produced by different runs will not be identical. To see how stable the algorithms are we run each method 20 times with different random seeds to generate partitions $P_i=\{c_{j,i}\}$, $i=1,\ldots,20$, and study the stability of the number of communities found ($|P_i|$), the size of the largest community ($S_{\text{max}} = \max_j\{|c_{j,i}|\}$) and the stability of identified communities across the runs. Let $\mathcal{P} = \{P_1,P_2,\ldots\}$ be a set of partitions and denote by $C_{\text{pm}}(\mathcal{P}) = \cap_{P \in \mathcal{P}}P$ the set of communities that appear in all partitions, i.e. the set of perfectly matching communities. For any $P_i \in \mathcal{P}$ the fraction of perfect matches is $f_{\text{pm}}(P_i;\mathcal{P}) = |C_{\text{pm}}|/|P_i|$. We denote by $f_{\text{pm}}^{\text{pair}}$ the fraction of perfect matches when $\mathcal{P}$ consists of two partitions, and by $f_{\text{pm}}^{\text{all}}$ the fraction of perfect matches when $\mathcal{P}$ consists of all 20 partitions generated by a single method.

\begin{table} \centering
\caption{Comparison of the stability of stochastic algorithms. We  generated 20 partitions with each method using different random  seeds, and present the smallest and largest observed values of   $|P_i|$ and $S_{\text{max}}$ over all 20 runs and of   $f_{\text{pm}}^{\text{pair}}$ over the 20 ordered pairs $(P_i,P_j)$ with $|i-j|=1$. The value of   $f_{\text{pm}}^{\text{all}} =   |C_{\text{pm}}(\{P_i\}_{i=1}^{20})|/|P_j|$ depends on the partition   only through $|P_j|$ and is therefore also very stable; we list the value corresponding to the largest $|P_j|$.}
\label{table:stability_test}
\begin{tabular}{c|r@{ -- }l|r@{ -- }l|r@{ -- }l|c}
 & \multicolumn{2}{c}{$|P_i|$} & \multicolumn{2}{|c}{$S_{\text{max}}$} & \multicolumn{2}{|c|}{$f_{\text{pm}}^{\text{pair}}$} & $f_{\text{pm}}^{\text{all}}$\\
\hline
IM & $280000$ & $280516$ & $2964$ & $3672$ & $42.1$ & $42.6$ \% & $13.2$ \% \\
LV & $1293903$ & $1298256$ & $811$ & $11390$ & $72.1$ & $72.8$ \% & $36.7$ \% \\
wIM & $674587$ & $674727$ & $209$ & $247$ & $97.4$ & $97.5$ \% & $92.5$ \%\\
wLV & $1155557$ & $1155985$ & $73$ & $112$ & $95.8$ & $95.9$ \% & $90.1$ \%\\
\end{tabular}
\end{table}

The results are summarized in Table \ref{table:stability_test}. It turns out that both weighted methods are very stable not only with respect to $|P_i|$ and $S_{\text{max}}$, but also with respect to the identity of communities: with both wIM and wLV we get $f_{\text{pm}}^{\text{all}} > 0.9$, which means that over 90 \% of communities are identical in all 20 runs. The variation comes mostly from large communities.

In the unweighted case both IM and LV are stable with respect to $|P_i|$, and IM also with respect to $S_{\text{max}}$. The identity of communities found however exhibits more variation: e.g., only 13 \% of communities found by a single run of IM appear in all 20 runs. Furthermore, looking at the unmatched communities for any pair (i.e. those in $P_i\backslash C_{\text{pm}}(\{P_i,P_j\})$), in IM about 32 \% have tiling imperfection $\mathcal{I} < 0.2$, and the average tiling imperfection is 0.46; in LV only 17 \% of such communities have $\mathcal{I} < 0.2$, with average tiling imperfection of 0.57. Thus the remaining communities are in general not even close matches. As with weighted methods, small communities are more likely to match perfectly than larger ones.

Instability of a method is of course problematic for anyone wanting to identify the ``true'' communities of a given network. It is however premature to judge IM and LV because of this: the network topology is inherently noisy, and does not necessarily contain enough information to uniquely identify the communities. Including weights made both methods much more stable, which suggests that the link weights contain information beyond the network topology. Note that there is information even in the instability: any two IM partitions share 42 \% of their communities, but if these shared communities were chosen uniformly at random only $0.42^{20} \approx 10^{-6}$ \% of the communities would appear in all 20 partitions---much less that the actual value of 13.2 \%.

The high stability of wIM and wLV may be partly explained by the fat-tailed distribution of call lengths in a mobile call network \cite{Onnela}. Since both methods are based on using probabilities proportional to the edge weights, an edge with a weight several orders of magnitude larger than the average will be placed inside a community almost independently of the network topology. On the other hand, in wCP the definition of intensity as the geometric average takes well into account the fat-tailed degree destribution, and is equivalent to using weights $w_{ij}^{\ast} = \log w_{ij}$, the arithmetic mean for intensity and the intensity threshold $I_>^{\ast} = \log I_>$. While one could use logarithmic weights also with wIM and wLV, this is problematic as the ratio of log-weights is not scale invariant and therefore the result would depend on the unit used to measure call length.

Finally, as suggested by the stability of $|P_i|$ and $S_{\text{max}}$, the qualitative properties of the communities are very stable even though the exact identity of communities are not. For example IM repeatedly produces treelike communities even if the communities are not made up of the same nodes. Because of this statistical stability no error is made by comparing the methods by using only single realizations from each method.

\end{document}